# Synthetic Low-Field MRI Super-Resolution Via Nested U-Net Architecture

Aryan Kalluvila[1], Danyal Bhutto[1,2,3], Neha Koonjoo[1,2], Marcio Rockenbach[2], Matthew S. Rosen[1,2,4]

[1]Athinoula A. Martinos Center for Biomedical Imaging, Department of Radiology, Massachusetts General Hospital, Charlestown, MA, United States, [2] Harvard Medical School, Boston, MA, United States, [3]Boston University, Boston, MA, United States, [4]Department of Physics, Harvard University, Cambridge, MA, United States

*Abstract*—Low-field (LF) MRI scanners have the power to revolutionize medical imaging by providing a portable and cheaper alternative to high-field MRI scanners. However, such scanners are usually significantly noisier and lower quality than their high-field counterparts. The aim of this paper is to improve the SNR and overall image quality of low-field MRI scans to improve diagnostic capability. To address this issue, we propose a Nested U-Net neural network architecture super-resolution algorithm that outperforms previously suggested deep learning methods with an average PSNR of 78.83 ± 0.01 and SSIM of 0.9551 ± 0.01. We tested our network on artificial noisy downsampled synthetic data from a major T1 weighted MRI image dataset called the T1-mix dataset. One board-certified radiologist scored 25 images on the Likert scale (1-5) assessing overall image quality, anatomical structure, and diagnostic confidence across our architecture and other published works (SR DenseNet, Generator RRDB Block, SRCNN, etc.). We also introduce a new type of loss function called natural log mean squared error (NLMSE). In conclusion, we present a more accurate deep learning method for single image super-resolution applied to synthetic low-field MRI via a Nested U-Net architecture.

I. INTRODUCTION

Magnetic resonance imaging (MRI) has revolutionized healthcare by providing a non-invasive diagnostic tool that can output high-resolution images of various anatomical structures. Although traditional MRI scanners operating at high magnetic field strengths (1.5T-3T) provide high sub-millimeter resolution scans, it is very cost-prohibitive and time-intensive due to the high magnetic field strength, installation costs, and operation of the machinery. As of 2021, there are approximately seven high-field MRI scanners per million inhabitants, and over 90% of such scanners are concentrated in high-income countries [1]. In 2020, the FDA approved the world's first portable low-field (LF) human MRI scanner that operates at a significantly lower magnetic field strength (of 64 mT) - the Hyperfine® scanner. Apart from being the most affordable scanners worldwide, specially to developing countries (around $50K/scanner) [2], LF scanners have enabled ICU patient scanning at bedside while being surrounded by ventilators and other metallic medical devices. However, lowering the magnetic field strength comes with its own challenges. Mainly, there is a significant drop in the signal-to-noise ratio (SNR) and overall image quality. In this paper, we focus on increasing the SNR and resolution of synthetic LF MRI scans via a Nested U-Net deep learning architecture. This neural network architecture has the base U-Net as the blueprint but utilizes redesigned skip connections [3].

While high-field (HF) MRI scanners provide high resolution (HR) to display anatomical structures, which is a requisite for diagnosis of many pathologies such as multiple sclerosis, smaller brain injuries, and neurocognitive diseases (Alzheimer's, Parkinson's). , there are medical specialties like emergency medicine where high resolution scans are not an immediate necessity and are therefore not cost-effective for local hospitals and clinics. Instead, a portable LF MRI scanner can be more appealing to such institutions provided the scanners output high enough spatial resolution and with sufficient diagnostic capabilities. One option for approaching this is improving the reconstruction method from the k-space to the image domain, however, this has only limited SNR improvement [4]. Another option would be to increase the low-field MRI acquisition time and magnetic strength; however, this makes the scan longer for the patient and makes it less suitable for portable bedside imaging [5].

Among the super-resolution (SR) deep learning approaches used to improve spatial resolution of images, single image super-resolution (SISR) has the capability to improve the SNR and overall image quality without altering any physical MRI properties. Previously, substantial work has been done on 3D low-field MRI super-resolution (or multi-image super-resolution (MISR)) [6]. In this paper, we focus on SISR because it requires less computational power and usually reports higher accuracy (due to greater dataset availability). Furthermore, very often radiologists analyze 2D MRI images to make a diagnosis, especially in emergency situations. 3D scans, even though they provide more information, take a great amount of time to analyze. In that way, 2D LF MRI SISR is more appropriate for emergency care imaging.

Current state-of-the-art SR methods involve three types of methods: interpolation-based, reconstruction-based, and learning-based techniques with the first two being analytical reconstruction methods, and the last one being a machine learning based approach [7-9]. Bilinear/bicubic interpolation techniques, while very computationally efficient, tend to over-smooth and provide granulated outputs, and tend to report small SNR improvements [10]. Reconstruction-based methods solve the blurred edges and granulation produced by interpolation by using a gradient and spatial extraction approach but lack finer details [8]. Learning-based techniques often utilize machine learning to bridge the resolution gap. These techniques report the highest SNR and overall image quality because they usually learn from a large dataset of paired degraded and HR images [9]. Andrew et al. reported using a lightweight autoencoder that leverages skip connections to sufficiently super-resolve downsampled high resolution MRI data [11]. M.L. de Leeuw den Bouter et al. trained an SR DenseNet and were able to inference on a low-field MRI scan to bridge the resolution gap [12]. Laguna et al. implemented a pipeline involving a domain adaptation network, a denoiser, and an SR block to adequately super-resolve 3D MRI scans [13]. The authors' network involved a Residual Dense Block in the ESRGAN generator to reconstruct LF MR images. Their work is focused on the domain adaptation portion of the network as bridging the gap between the low-field image domain and the high-field image domain, which is quite difficult. For simplicity, we assume domain adaptation will be sufficiently close to high-field MRI data so that our SR block can reconstruct properly.

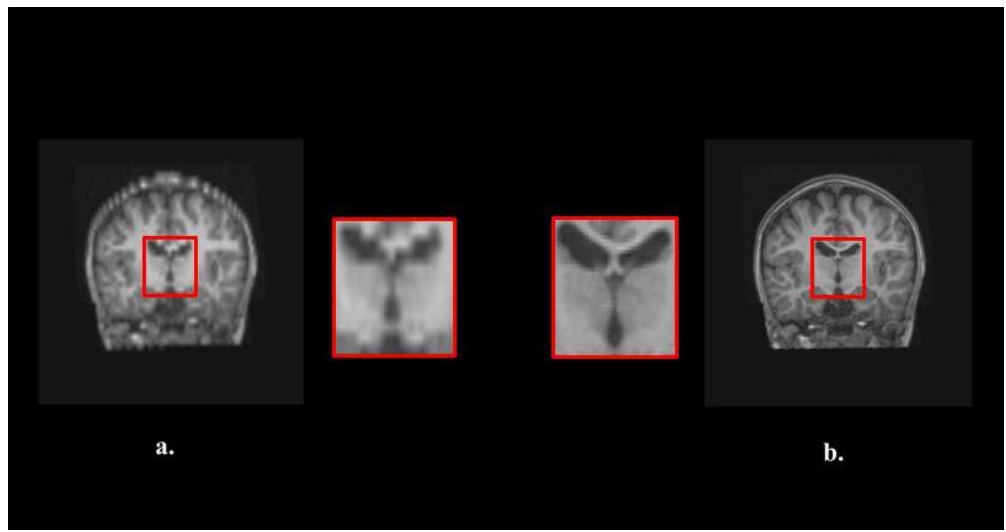

Figure 1: Sagittal slices of synthetic 64 mT MRI scans (1.) and 3T MRI counterparts (2.) The 3T scans have much higher resolution, greater anatomical structure, and changed contrast.

In Figure 1a, the structural integrity of the inner regions are significantly distorted (which reflected low-field MRI quality). In Figure 1b, the ground truth scan retains this integrity and heightened contrast. In this paper, we attempt to bridge the gap between these two scans in terms of both contrast and overall quality.

In this paper, we propose using a SR U-Net++ architecture (Nested U-Net) to reconstruct HR images from synthetically downsampled LF images. We trained the network to output the difference between the high-field and synthetic low-field data as shown in Figure 2. We compared our technique to state-of-the-art methods that included SRCNN, VDSR, and a variation of the SRGAN generator. The major contributions of this work are as follows:

- A down sampling pipeline that accurately transfers HF MRI images to LF MRI images.
- A trained U-Net++ architecture for synthetic LF MRI scans to achieve SISR through employing residual learning
- Comparison of the U-Net++ architecture against state-of-the-art algorithms to evaluate performance through PSNR, SSIM, and a board-certified radiologist

II. MATERIALS AND METHODS

*A. Dataset Preparation*

In this study, a total of 1,500 T1-weighted 3T MR human brain images were used. The datasets were primarily the 1,500 3T scans from the T1-mix dataset, that was classified as Dataset I. All the scans in this dataset were in coronal view. Dataset I contained only autism scans but no pathological lesions were included, contributing to the robustness of our U-Net++ trained model [20].

All 1,500 MRI scans were acquired using a GE 3T MR750 scanner with an 8-channel head coil at the UCSD Center for Functional MRI [14 & 20]. The HR scan sequences were acquired from FSPGR T1-weighted sequence (TR: 11.08ms; TE: 4.3ms; flip angle: 45°; FOV: 256mm; 256 x 256 matrix;
180 slices; 1mm3 in-plane resolution) [15]. Foam pillows were put around the patients' heads to minimize movement [15].

*B. Implementation Details*

The Database I was split into 3:1 for the training set and the validation set. The input to the architecture were the synthetic 64 mT LF MRI scans that were generated using a downsampling pipeline from the ground truth HF scans, as described below in the Methods section C. The ground truth were the HF 3T MRI scans (256 x 256) and the residual image (Synthetic LF image – HF image) were used as the corresponding target / output of the architecture. The main SR algorithm was based on our proposed U-Net++ with VGG blocks and without batch normalization (as described in Methods sections D and E). Both input and output images had only one channel (not RGB) to increase computational efficiency. Six different state-of-the-art algorithms - VDSR, SR-CNN, SR-GAN, DenseNet, SR U-Net, U-Net++ were trained with the same parameters as specified below and evaluated.

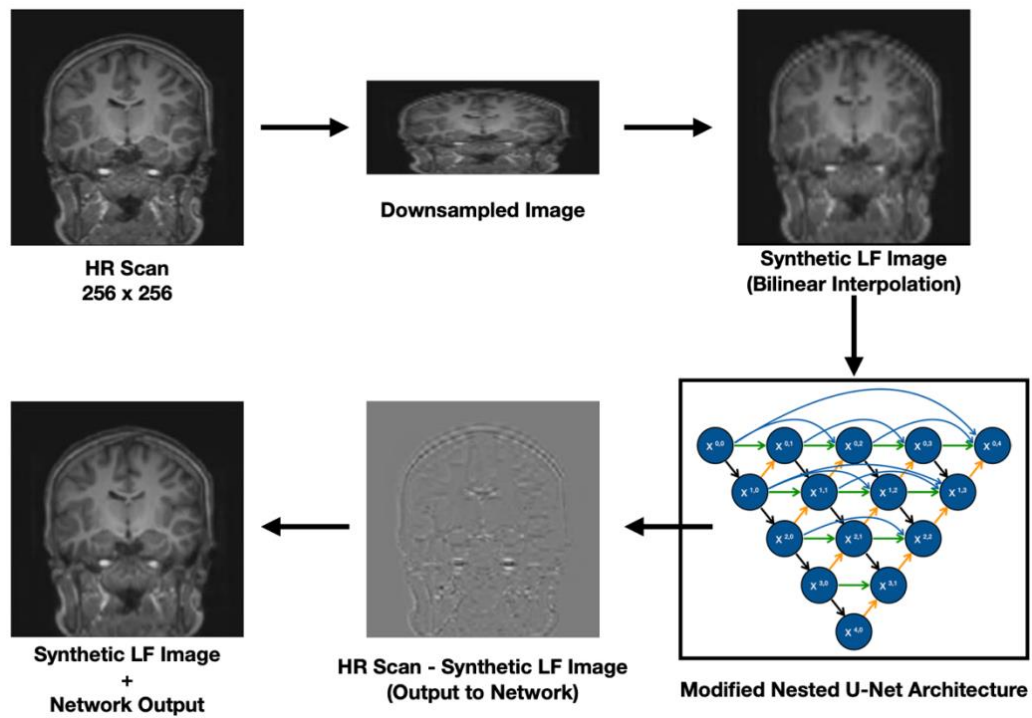

Figure 2: The downsampling pipeline to produce synthetic LF MRI. It begins with 256 x 256 HR images and is corrupted through downsampling and bilinear interpolation and is brought to back up to resolution by U-Net++.

The network was trained with the Adam optimizer with a learning rate of $1e^{-4}$. Weight decay or L2 regularization was also employed in the network with a value of $5e^{-4}$ which helped prevent the network from overfitting. The entire pipeline was trained using PyTorch 1.9.2 with a NVIDIA Quadro RTX 6000 GPU and 24 GB RAM.

All 6 neural networks were trained for 60,000 epochs with a batch size of 1 (no batch normalization was employed for consistency and comparison purposes). Each network's performance was evaluated using the image metrics PSNR and SSIM.

*C. Downsampling Pipeline*

Data collection for training an SR network usually requires paired HF and LF scans of the same patient. However, this method involves non-linear distortions, as well as patient registration for a perfect match. Here, to attain LF MRI quality from HF MRI scans, a unique and efficient downsampling approach was used as described in Figure 2. Usually, scans are downsampled by the same factor vertically and horizontally. However, at LF, the images are distorted in a unique way. In an empirical manner, applying an asymmetric downsampling factor in the horizontal and vertical directions of 1.5 and 5 respectively resulted in a less distorted downsampled image compared to a symmetrical downsampled image. This elongated downsampled MRI scan of 52 pixels × 172 pixels was then rescaled using the bilinear interpolation to 256 pixels × 256 pixels resolution. This final interpolated image was the input to the U-Net++ architecture. This process produced a cleaner and more efficient downsampling method and training pipeline.

To improve the performance of the training, all images of the Dataset I were normalized from -0.5 to +0.5. Rigorous data augmentation was performed like random affine, blur, crop, random rotation, etc. thus improving the robustness of the neural network and reducing overfitting in the training process. Furthermore, to reduce the computational time, we used a technique called residual learning. As shown in Figure 2 the output/target of the neural network was the difference between the 256 × 256 HF image and the 256 × 256 bilinear interpolated LF image. To reproduce the final SR image, the output was then overlaid onto the bilinear interpolated LF image. This residual learning approach improved the performance of the training since less pixel information is required to be learned by the U-Net++ architecture.

*D. Network Architecture*

This section introduces the U-Net++ architecture which is the primary super-resolution algorithm used to reconstruct our synthetic LF brain images. As a well-known architecture in the literature, the standard U-Net involves a symmetric "U" shaped architecture that has a contrastive and expansive path (similar to the variational autoencoder) [16]. Each downsampling and upsampling layer involves two convolutional layers with padding. The bottom layer (bottleneck) also has two convolutional layers, but no max pooling. Transposed convolutions are used to upsample from the bottleneck layer to the final image size. The benefit of using a U-Net over other architectures is that it is able to localize and distinguish borders better (due to the classification on each pixel layer) which is suitable for biomedical applications (such as segmentation tasks). Like the U-Net, the U-Net++ described by Zhou Z et al, involved an encoder and decode to bridge the semantic gap between feature maps prior to diffusion [17]. As shown in Figure 3, the standard U-Net is represented in black circles and the additional skip connections are represented in green and blue dotted arrows. The main distinction between the two networks was the redesigned skip connections (denoted in green and blue dotted arrows). In the standard U-Net, the feature maps of the encoder are directly fed into the decoder, and, our U-Net++ were composed with multiple dense convolutions to improve accuracy of the feature maps before being fed into the decoder. For instance, the skip connection pathway between $X^{1,0}$ to $X^{1,3}$ is composed of three convolutional layers with one dense block where each convolutional layer is preceded by a concatenation layer that fuses the output from the previous convolutional layer [17]. The dense convolutional layers bring the encoder feature maps closer to the decoder feature maps so the accuracy of the overall network is improved.

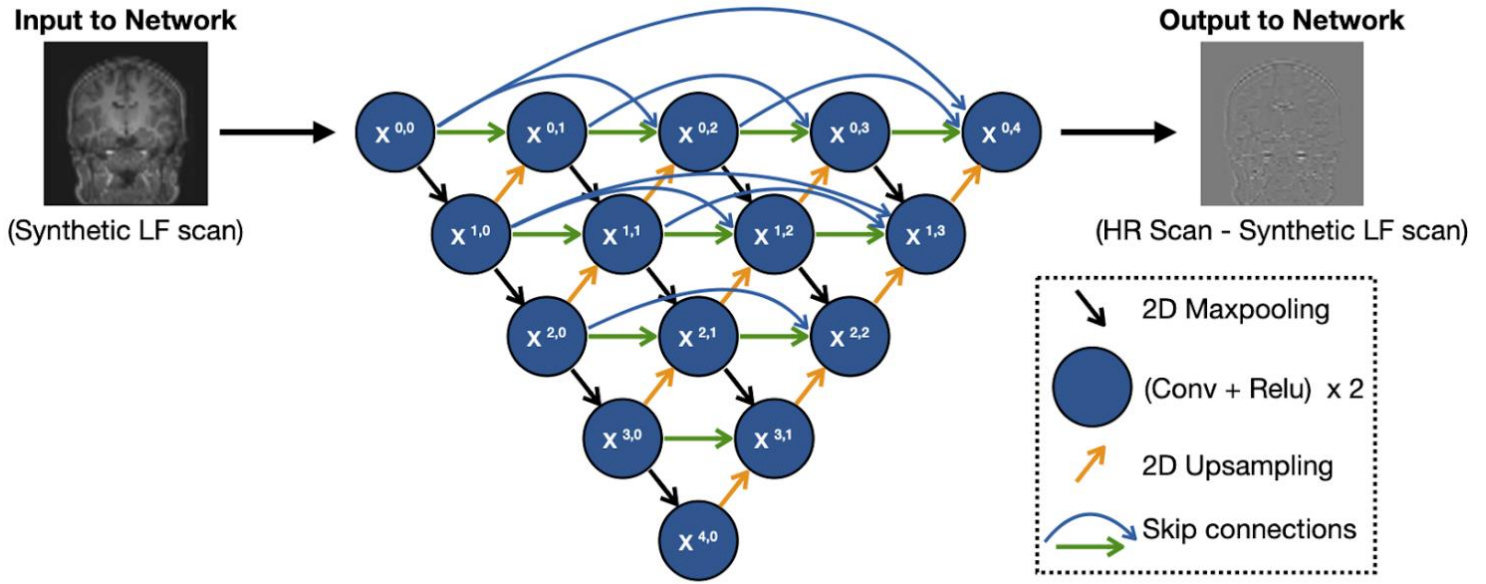

Figure 3: Our Proposed U-Net++ architecture for SR algorithm. The blue circles represent the standard U-Net architecture and the colored blue and green symbols represent the added convolutioinal layers and skip connections of the U-Net++ architecture.

The skip pathway was created as follows: let $x^{i,j}$ denote the output node $X^{i,j}$ where $i$ indexes the down-sampling layer along the encoder and $j$ indexes the convolutional layer of the dense block along the skip pathway [17]. Z. Zhou et al. also proposed deep supervision on top of the U-Net++ architecture, however, there was no apparent benefit in using deep supervision as that was primarily for segmentation purposes.

Instead of using single traditional convolutional layers, visual geometry group blocks (VGGs) were used. VGG blocks are generally composed of multiple convolutional and max pooling layers. Traditional U-Net++ architectures with VGG consist of 1 ReLU activation unit, 2 convolutional layers, and 2 batch normalization layers. Hu et al. proposed a U-Net for image super-resolution without batch normalization and reported a significant increase in overall image quality and resolution [18]. Hence, batch normalization was not used in our proposed U-Net++ architecture as well.

*E. Loss Function Approximation*

In most super-resolution image reconstruction tasks, the loss function employed is the mean squared error loss function as shown in Equation 1,

$$MSE = \frac{1}{n}\Sigma(Y_i - Y_j)^2 \quad (1)$$

where *n* represents total image size, $Y_i$ represents the output of the network, and $Y_j$ represents the residual target (synthetic LF – HF). Minimizing such errors produced satisfactory results for random natural images (such as dog, cat, person), but it failed to reconstruct the precise anatomical structures in MRI and CT scans [19]. In this study, we experimented with several different loss functions including: MSE, VGG, and abs-MSE. All of these loss functions produced grainy results. Thus, we proposed a new SR loss function that outperformed the previous loss functions called natural log mean squared error loss (NLMSE), as stated with Equation 2:

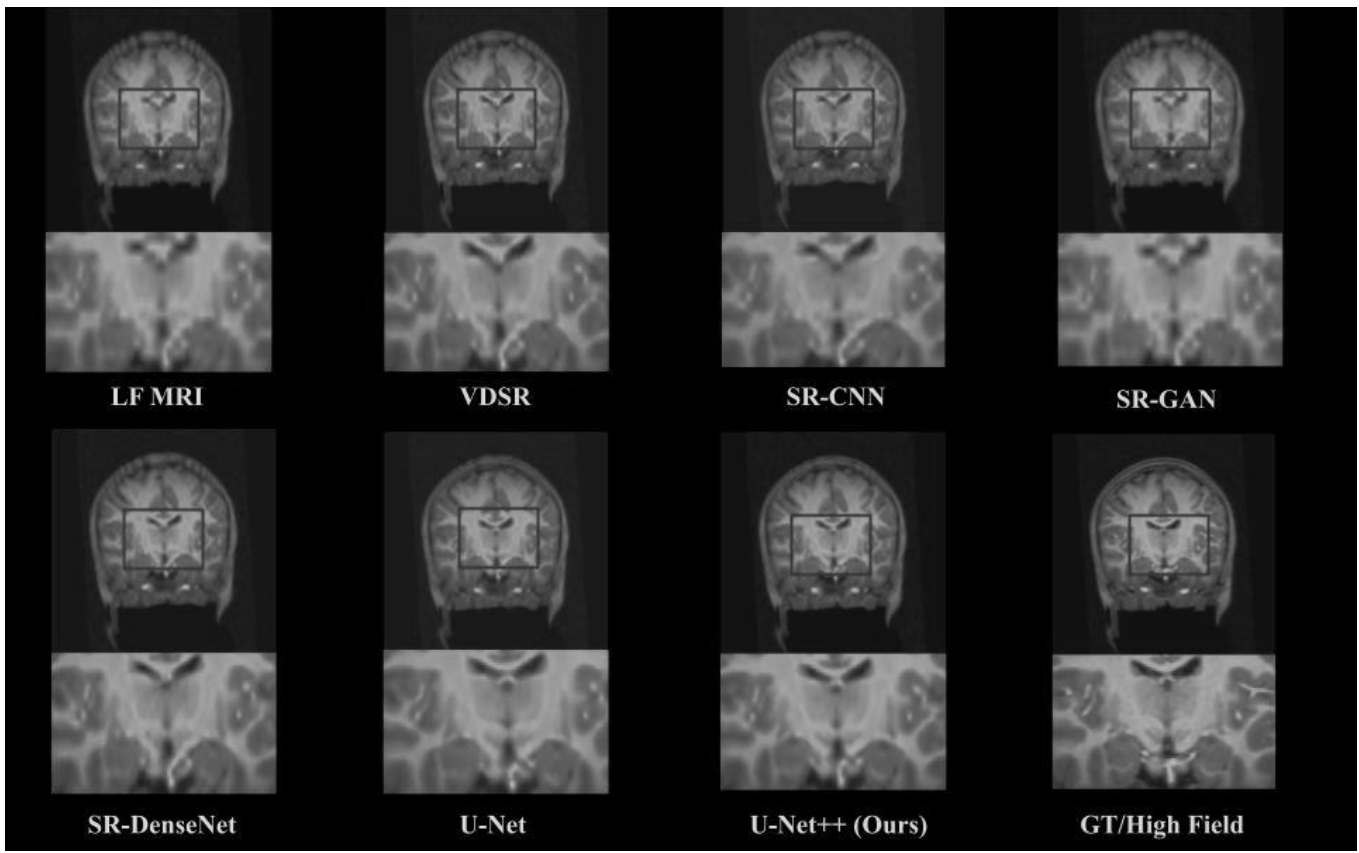

Figure 4: Qualitative observations of LF MRI scans vs super-resolved scans via 5 other networks and the U-Net++ for comparison. The GT/High Field being the reference point.

$$NLMSE = \frac{1}{n} log_e(\sum(Y_i - Y_j)^2) \quad (2)$$

## III. Model's Performance and Comparison

### A. Comparison Against State of the Art Algorithms

To rigorously test the proposed SR U-Net++ architecture, we compared the algorithm against five other state of the art networks which include VDSR, SR-CNN, SR-GAN, DenseNet, SR U-Net. We evaluated all six networks using PSNR and SSIM to get a quantitative evaluation of our networks.

We tried to limit our state-of-the-art architectures to SR algorithms that have been applied to MRI super-resolution. Table 2 shows the comparison against state-of-the-art algorithms and the U-Net++ outperforms all networks in terms of both PSNR and SSIM. Comparable to other networks, there isn't marginal improvement. The performance of our proposed network demonstrated the largest increase in PSNR and SSIM performance (+0.29 in PSNR/+0.124 in SSIM).

*Table 1: PSNR and SSIM against state-of-the-art algorithms applied to Dataset I.*

| Dataset I | PSNR (db) | SSIM |
|---|---|---|
| LF MRI | 75.52 | 0.9159 |
| VDSR | 78.10 | 0.9481 |
| SRCNN | 77.56 | 0.9427 |
| SR GAN | 75.52 | 0.9159 |
| SR U-Net | 78.54 | 0.9519 |
| SR DenseNet | 78.06 | 0.9477 |
| **SR U-Net++** | **78.83** | **0.9551** |

### B. Mean Opinion Score (MOS) Testing

A board-certified radiologist (M.R.) was given 5 scans per network (30 total images) to assess. This was a blinded study where the radiologist had no information on which neural network was used to reconstruct to prevent any bias. The radiologist was asked to rate each subsequent image on a scale from 1-5, with 1 being very close to LF and 5 being very close to HF (the Likert scale). Initially, the radiologist was calibrated by observing 5 LF MRI scans and 5 HF MRI scans to learn the scale. They weren't allowed to rank in between numbers (such as 3.5). These scores were then averaged and written in Table 3.

The following calculations are averaged between the radiologist's evaluation of all five images.

*Table 2: Mean Opinion Score Evaluation from Radiologist on State of the Art Algorithms*

| Dataset I | Mean Opinion Score |
| --- | --- |
| LR MRI | 1.0 ± 0.0 |
| VDSR | 3 ± 0.70 |
| SRCNN | 3.2 ± 0.44 |
| SR GAN Generator | 1.0 ± 0.0 |
| SR U-Net | 3.6 ± 0.55 |
| SR DenseNet | 3.4 ± 0.89 |
| **SR U-Net++** | **4.4 ± 0.54** |

The mean opinion score study corroborated well with our PSNR and SSIM studies. In the blind study, the radiologist chose the U-Net++ architecture as the highest performing network. The U-Net++ outperformed all other networks by at least 0.3 MOS. The radiologist noted improved reconstruction in the hippocampal and skull regions where other networks exhibited artifacting in those regions such as the zebra stripe pattern and bigger hippocampal volume. This study was completed to verify that the U-Net++ reconstructed clinically relevant details. The second best performance was the SR DenseNet proposed by M.L. de Leeuw den Bouter.

## IV. DISCUSSION AND CONCLUSION

In this study, we demonstrated accurate synthetic low-field MRI super-resolution to 3T MRI using a U-Net++ architecture. Our network outperformed current state-of-the-art networks by a substantial amount. This work demonstrates the promise of fully connected U-Nets for medical image super-resolution tasks, especially when filling a larger resolution gap. Previous MRI super-resolution papers aimed to bridge a smaller resolution gap which marginally improves the scanner. However, in this study, we show that a U-Net++ can substantially improve the anatomical resolution of MRI scans with high PSNR and SSIM values.

From Figure 7, there is a substantial improvement in resolution from the other networks and the SR U-Net and SR U-Net++, especially in the inner regions of the brain. The difference between the SR U-Net and the SR U-Net++ is the slight contrast improvement. The redesigned skip connections allow for greater improvement within those areas where the traditional SR U-Net fails. In Table 1, the PSNR and SSIM of the SR U-Net++ outperforms all other state of the art. The U-Net comes in a close second but again hinders due to the contrast difference. A similar SR-DenseNet architecture proposed by M.L. de Leeuw den Bouter was also tested in this study which closed around a 78.08 PSNR and 0.9477 [12]. From Figure 7, this network exhibits some artifacts in the two brain cavities magnified. U-Nets specialize in localization which enables them to smooth out these regions and obtain substantially higher PSNR and SSIM values.

This study has only been tested on synthetic LF MRI data. Laguna et al. emphasized the importance of domain adaptation in realistic LF MRI super-resolution as the imaging sequences cause the domain shift to be quite substantial [13]. Additionally, we didn't include add any Gaussian noise distributions to any of our scans to mimic LF MRI conditions. This is another critical portion of achieving stronger LF MRI super-resolution results. We focus primarily on bridging the resolution gap after the domain adaptation and denoiser have been successfully implemented. Using CycleGANs to create synthetic LF MRI scans could also be show some improvement in realistic LF MRI scenarios. Transformers have seen recent news in NLP and visual machine learning tasks. Applying such algorithms to medical image super-resolution could also yield improvement in results. Future studies could involve concatenation of a strong domain adaptation network, a denoiser, and an SR block (like the one we proposed in this study) to corroborate strong LF MRI super-resolution directly from the scanner itself. Here, we seek to improve one part of that pipeline. Also, although we have extensively applied the SR U-Net++ to brain MRI super-resolution, applying it to other organs at LF (especially cardiac imaging) could show the robustness of the SR U-Net++ even further. Additionally, in machine learning terms, we had a relatively small dataset of approximately 2,000 images. Increasing this dataset and performing rigorous augmentation could potentially improve results. Finally, we hope to evaluate our network on actual LF MRI scans in the future as well.

In this paper, we propose a SR U-Net++, previously used for medical image segmentation, to the task of medical image super-resolution. Specifically, we apply this to LF MRI SISR. From a dataset of about 2,000 images taken from healthy patients and autistic patients, we create synthetic LF MRI images using a unique downsampling pipeline designed for 64 LF MRI reconstruction. For the primary reconstruction pipeline, we used a U-Net++ which takes on the original U-Net architecture but redesigns the skip connections instead of just directly feeding the features maps from the encoder to the decoder. The skip connections are VGG convolutional blocks stripped of batch normalization to improve decoder accuracy. We also propose a new SR loss function called NLMSE which improves accuracy substantially. From PSNR and SSIM studies, the U-Net++ outperforms all tested state-of-the-art algorithms. For the qualitative inspection, the U-Net and U-Net++ recover local pixel details at greater detail than any other tested network. The U-Net++ is speculated to have improved contrast, however with a PSNR of 78.83 and SSIM and 0.9551. We also completed MOS testing to verify the clinic relevance of the reconstructed areas from the U-Net++. In this blind study, the U-Net++ was chosen as the highest performing network. Overall, the U-Net++ is a strong contender for LF MRI super-resolution. In the future, we would like to apply our algorithm to clinical LF MRI scans to see how it would work after it has been sufficiently denoised and domain adapted to 3T MRI. We also foresee our proposed network being used in other organ imaging applications.